\documentclass[galaxies,article,submit,moreauthors,pdftex,10pt,a4paper]{mdpi}

\firstpage{1}
\articlenumber{x}
\doinum{10.3390/------}
\pubvolume{xx}
\pubyear{2017}
\copyrightyear{2017}
\externaleditor{Academic Editor: Emilio Elizalde}
\history{Received: 13 September 2017; Accepted: 13 October 2017; Published: date}

\usepackage{booktabs}
\usepackage{multirow}
\usepackage{soul}
\usepackage{microtype}
\usepackage{upgreek}
\Title{3 mm GMVA Observations of Total and Polarized Emission from Blazar and Radio Galaxy Core Regions}

\Author{{Carolina Casadio} %
$^{1}$*, Thomas P. Krichbaum $^{1}$, Alan P. Marscher $^{2}$, Svetlana G. Jorstad $^{2,3}$, Jos\'e~L.~G\'omez $^{4}$, Iv\'an Agudo $^{4}$, Uwe Bach $^{1}$, Jae-Young Kim $^{1}$, Jeffrey~A.~Hodgson $^{5}$ and Anton~J.~Zensus $^{1}$}

\AuthorNames{Carolina Casadio, Thomas P. Krichbaum, Alan P. Marscher, Svetlana G. Jorstad, Jos\'e L. G\'omez, Iv\'an Agudo, Uwe Bach, Anton J. Zensus.}

\address{%
$^{1}$ \quad Max-Planck-Institut f\"ur Radioastronomie, Auf dem H\"ugel, 69, D-53121 Bonn, Germany\\
$^{2}$ \quad Institute for Astrophysical Research, Boston University, Boston, MA 02215, USA\\
$^{3}$\quad Astronomical Institute, St. Petersburg State University, St. Petersburg 199034,
Russia\\
$^{4}$ \quad Instituto de Astrof\'{\i}sica de Andaluc\'{\i}a, CSIC, Apartado 3004, E-18080 Granada,
Spain\\
$^{5}$ \quad Korea Astronomy and Space Institute, 776 Daedeokdae-ro, Yuseong-gu, Daejeon 34055, Korea}

\corres{Correspondence: casadio@mpifr-bonn.mpg.de} %Please add correspondence author.

\abstract{We present total and linearly polarized 3 mm Global mm-VLBI Array (GMVA; mm-VLBI: Very Long Baseline Interferometry observations at millimetre wavelengths) images of a sample of blazars and radio galaxies from the VLBA-BU-BLAZAR 7 mm monitoring program designed to probe the innermost regions of active galactic nuclei (AGN) jets and locate the sites of gamma-ray emission observed by the Fermi-LAT. The lower opacity at 3 mm and improved angular resolution---on the order of 50 microarcseconds---allow us to~distinguish features in the jet not visible in the 7 mm VLBA
~data. We also compare two different methods used for the calibration of instrumental polarisation and we analyze the resulting images for some of the sources in the sample.}


\begin{document}
\section{Introduction}

Combining long baselines and short observing wavelength, Very Long Baseline Interferometry observations at millimetre wavelengths (mm-VLBI)  provide very high spatial resolution images. Moreover, the reduced opacity at 3 mm allows us to investigate regions that are optically
thick at longer wavelengths, such as the jet formation regions in the vicinity of supermassive
black holes. How~jets are formed, accelerated, and collimated in such regions is
still under debate, and high-resolution polarized images are the key to investigating the energy budget and magnetic field structure that, according to theoretical models (e.g., \citep{Tchekhovskoy:2011kq}), should play a major role in these processes.

The Global mm-VLBI Array (GMVA), observing at 3 mm (86 GHz), is currently the main source of high-resolution total and polarized intensity images of active galactic nuclei (AGN) at short millimeter wavelengths.
The other way to reach very high resolutions---but at lower frequencies---is to observe with the space VLBI technique, which has provided the highest angular resolution ($\sim$21 $\mu$as)
achieved to-date~\citep{Gomez:2016qy}.
In the near future, VLBI imaging at 1.3 mm will provide an even higher resolution than that which is currently possible with the GMVA.

The GMVA consists of 14 antennas located in Europe, corresponding to Effelsberg (EF), Onsala~(ON), Pico Veleta (PV), Plateau de Bure (PdB), Mets\"ahovi (MH) and Yebes (YS), and in the United States, including the Green Bank Telescope (GBT) and the eight VLBA stations equipped with 3 mm receivers (BR, NL, PT, LA, FD, KP, OV, MK). Korean stations (KT, KU and KY) have also recently joined some GMVA sessions.

Here we present some 86 GHz GMVA images in both total and linearly polarized intensity of a~sample of bright and gamma-ray loud blazars taken on 21 May 2016 within a program (PI:~A.~Marscher)
in support of the VLBA-BU-{BLAZAR} \footnote{\url{http://www.bu.edu/blazars/VLBAproject.html}}
monitoring project; where the latter consists of monthly observations of 37 blazars and radio galaxies with the VLBA at 7 mm (43 GHz).
The GMVA program was started in 2008, consists of one~or~two observations per year of roughly half of the
AGN in the VLBA-BU-BLAZAR sample, and is aimed at relating the gamma-ray emission observed in these objects to physical conditions and structure in the mm-wave core region.
Information about the program as well as some preliminary results can be found
at \url{http://www.bu.edu/blazars/vlbi3mm/}, as well as in \cite{Hodgson:2015uk,Hodgson:2017yg}.

%%%%%%%%%%%%%%%%%%%%%%%%%%%%%%%%%%%%%%%%%%
\section{GMVA Observation and Data Reduction}

The antennas participating in our observations during the GMVA session of May 2016
were the VLBA stations plus Effelsberg, Onsala, Yebes, Mets\"ahovi, and the
KVN array, although it was not possible to recover fringes from Mets\"ahovi. All
antennas recorded the data at a rate of 2 Gbps (512~MHz bandwidth) in dual-polarization mode (i.e., right and left circular polarization, RCP and LCP, respectively), apart from Yebes that only observed in LCP; the data were divided into eight 32~MHz sub-bands (IFs) per polarization during the correlation process.

The observed sources, for which the calibration and imaging of both total and
linearly polarized intensity have been performed, are: 3C~111, 3C~120, 0716+714, OJ~287, 0954+658, 3C~273, {1510-089},
1633+382, 3C~345, BL~Lac, CTA~102, and 3C~454.3.

Data were fringe-fitted and calibrated using the common procedure for high-frequency VLBI data reduction in the Astronomical Image Processing System {\tt (AIPS)} (e.g.,
\citep{Jorstad:2005fk}), with the main difference of using the { manual phase-calibration}
approach instead of the so-called { phase-cal injection tones} for the
correction of delays and phases of the subbands as described in
\cite{Marti-Vidal:2012yq}.~The phase calibration is the trickiest part, and we found that remaining delays or
phase-offset between IFs can easily reduce the coherence of the signal.
Phase instabilities mainly come from instrumental and atmospheric effects; at~86~GHz, the troposphere has a very short expected
coherence time ($\sim$10--20 s). In addition and mainly due to the changing weather conditions, the accuracy of the
amplitude calibration is more limited at mm- than at cm-wavelength.

After the calibration of the right--left phase difference in {\tt AIPS}, we have to
correct for the instrumental polarization and calibrate the absolute orientation
of the polarization electric vector position angles (EVPAs).~The calibration of EVPAs requires the comparison between integrated GMVA EVPA measurements and single-dish EVPA values at the same observing frequency and at nearby epochs.
Hence, for the calibration of polarization vectors we used the information contained in 3~mm polarimetric measurements from the POLAMI program \footnote{see \cite{Agudo:2017b} and \url{http://polami.iaa.es}}, with uncertainties in the EVPAs $\leq$ 5$^{\circ}$ \citep{Agudo:2017a}.
For the calibration of instrumental polarization instead, we~tested two different approaches and finally used the most reliable one as we describe in Section \ref{pol}.

\subsection{Comparison of Total Flux Densities}
\label{tot_flux}
In order to check the reliability of the results obtained and in particular of the amplitude calibration, we compare the obtained total flux densities of each source with single dish data of nearby epochs
collected from the POLAMI program at the IRAM 30m Telescope (in Granada, Spain), at 3mm, and from the Sub-millimeter Array (SMA)
\footnote{\url{http://sma1.sma.hawaii.edu/callist/callist.html}}, at 1mm.
 %We also compare our images and flux measurements with the VLBA-BU-BLAZAR data on 10th of June 2016.
The fluxes from the three programs are reported in Table \ref{tab1}, where we notice that the flux
density values are in fairly close agreement between each other only for sources
like 0716+714, 0954+658, 1510-089, and 1633+382, that are core-dominated blazars \citep{Jorstad:2017qy}. For the other more extended sources, discrepancies between the GMVA and single dish values are present.
These are in part expected because of the weaker extended jets contributing more in single dish data, but we cannot discard a lowering of the flux in GMVA data with low signal-to-noise ratio.
That could be the case for weak sources like BL Lac, which in this observation was barely detected.

\begin{table}[H]
\centering
 \caption{This table reports the names of sources and total flux densities in mJy obtained for Global mm-VLBI Array (GMVA) data on 21 May 2016 and single dish data from the {POLAMI}
~(3 mm) and Sub-millimeter Array (SMA) (1 mm) programs. The~POLAMI observing epochs are 14 May (first row) and 14 June 2016 (second row), while the SMA epochs are variable but all separated by $\sim$15 days from the GMVA epoch, apart from 3C~273, 3C~454.3, and 0716+714, where the separation is less than 5 days. For GMVA data we estimate errors to be roughly~10$\%$ of the flux, as reported in \cite{Hodgson:2015uk}.}
%% \tablesize{} %% You can specify the fontsize here, e.g.  \tablesize{\footnotesize}. If commented out \small will be used.
\begin{tabular}{cccc}
\toprule
\textbf{Name}	& \textbf{GMVA}	& \textbf{POLAMI} & \textbf{SMA}\\ % & VLBA-BU\\
\midrule
3C~111		& 599 $\pm$ 60	& 1523 $\pm$ 84 & 991 $\pm$ 53\\ % & 1376$\pm$130\\
&& 1293 $\pm$ 57 &\\
\midrule
{3C~120}		& 726 $\pm$ 30 & 2627 $\pm$ 104 & 1970 $\pm$ 102\\ % & 2368$\pm$230\\
&& 2137 $\pm$ 97 &\\
\midrule
{3C~273}		& 7563 $\pm$ 760 & -- & 10,815 $\pm$ 541\\ % & 16999$\pm$1600\\
\midrule
{3C~345}		& 1582 $\pm$ 160 & 2300 $\pm$ 94 & 1689 $\pm$ 91\\ % & 3125$\pm$300\\
&& 2994$\pm$115 &\\
\midrule
{3C~454.3}	& 4178 $\pm$ 400 & 9814 $\pm$ 1509 & 10,100 $\pm$ 160\\ % & 14066$\pm$500\\
&& 12,070 $\pm$ 462 &\\
\midrule
{0716+714}	& 2455 $\pm$ 250 & 2076 $\pm$ 85 & 2132 $\pm$ 107\\ % & 2075$\pm$200\\
&& 2506 $\pm$ 102 &\\
\midrule
{0954+658}	& 572 $\pm$ 60 & -- & 786 $\pm$ 40\\ % & 905$\pm$90\\
\midrule
{1510-089}	& 2418 $\pm$ 240 & -- & 2664 $\pm$ 139\\ % & 4580$\pm$460\\
\midrule
{1633+382}	& 1027 $\pm$ 100 & 1362 $\pm$ 59 & -- \\ % 1816$\pm$180\\
&& 1977 $\pm$ 79 &\\
\midrule
{BL~Lac}	& 630 $\pm$ 60 & 1952 $\pm$ 75 & -- \\ % 1472$\pm$70\\
&& 1683 $\pm$ 60 &\\
\midrule
{CTA~102}	& 2736 $\pm$ 280 & 4815 $\pm$ 182 & 4429 $\pm$ 222\\ % & 3017$\pm$300\\
&& 4723 $\pm$ 180 &\\
\midrule
{OJ~287}	& 1767 $\pm$ 170 & -- & 2932 $\pm$ 159\\ % & 4580$\pm$350\\
\bottomrule
\label{tab1}
\end{tabular}
\end{table}

\section{Linear Polarization at 86 GHz: Instrumental Polarization Calibration}
\label{pol}

\begin{figure}[H]
\centering
\includegraphics[width=12 cm]{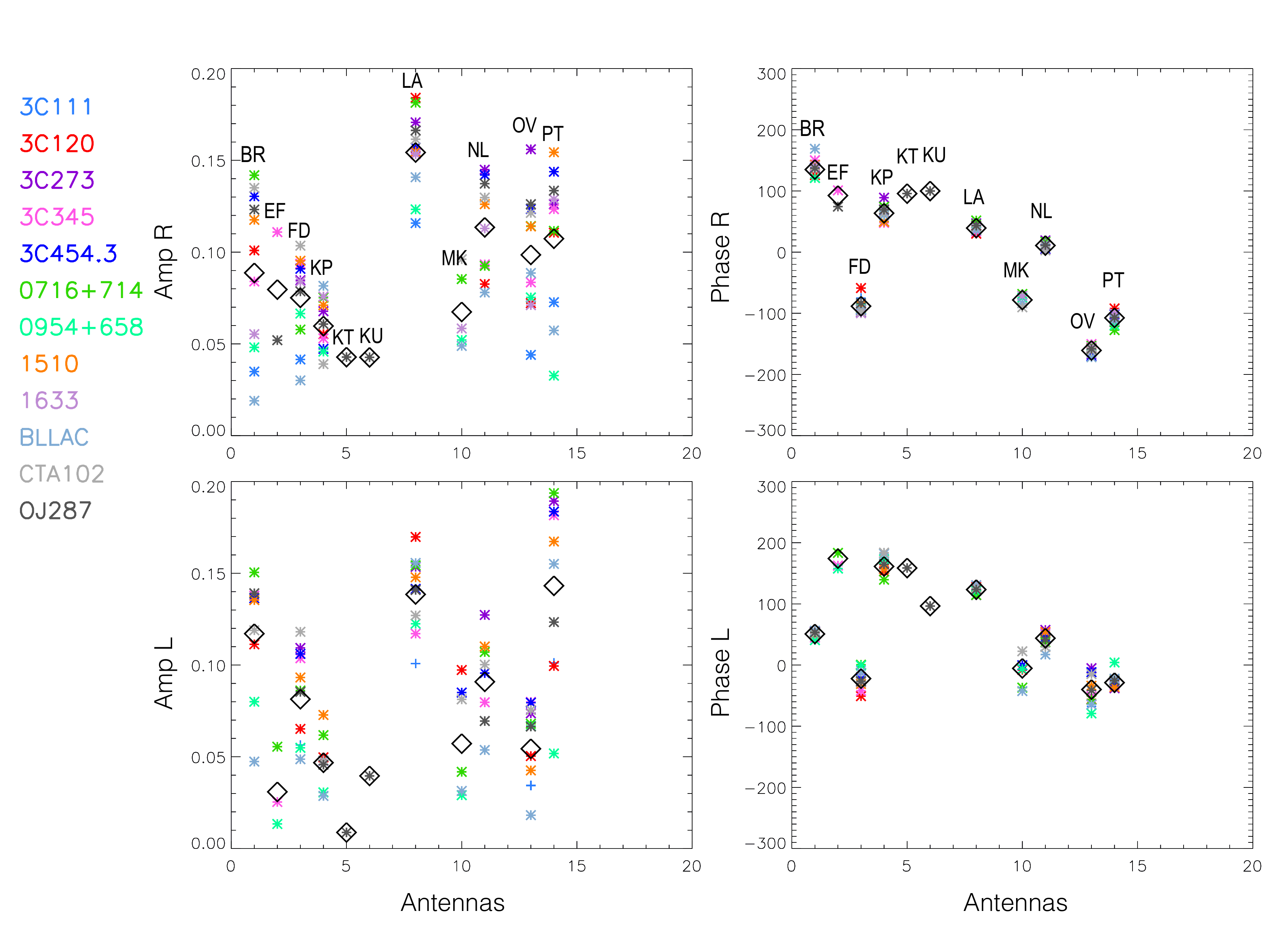}
\caption{The right and left amplitudes and phases of the {\textit D-terms} for all the sources in different colours. The codes of antennas (see text) are also reported inside the two upper plots. Black diamonds mark the average {\textit D-term} values obtained for each antenna, considering the values of the remaining sources after removing outliers (not displayed in the figure).}
\label{dterms}
\end{figure}

The instrumental polarization---or ``{\textit D-terms}''---were determined using the method of \cite{Leppanen:1995fv}. The~{\textit D-terms} are expected to depend only on the receiver characteristics of each telescope and to~change slowly with time. Hence, all the sources observed must possess the same {\textit D-terms}, although
the accuracy of the measurements depend mostly on the actual parallactic angle coverage,
and in part also on the complexity of the polarized source substructure.
Considering this, one possible approach for the calibration of the instrumental polarization is
to apply to each source a set of {\textit D-terms} that come from the average of the values obtained for all sources after removing outliers (e.g., \citep{Marscher:2002ys}). In Figure \ref{dterms} we report the {\textit D-term} amplitudes and phases of the RCP and LCP feed
for all the sources in the sample. The average {\textit D-terms} obtained are represented by black diamonds.

In the second approach tested, we imaged each source using the {\textit D-term} obtained from its own data.
Both methods have been tested previously in similar GMVA observations in \cite{Marti-Vidal:2012yq}, where the authors claim that no differences were observed between the two methods.

We tested both methods on OJ~287, which presents the best parallactic angle coverage (between 100$^{\circ}$ and 120$^{\circ}$ for most of the antennas) and has a simple and compact polarized
structure, and in two~other less favourable sources: 1510-089 and CTA~102.

%%%%%%%%%%%%%%%%%%%%%%%%%%%%%%%%%%%%%%%%%%
\section{Results}

\begin{figure}[H]
\centering
\includegraphics[width=13 cm]{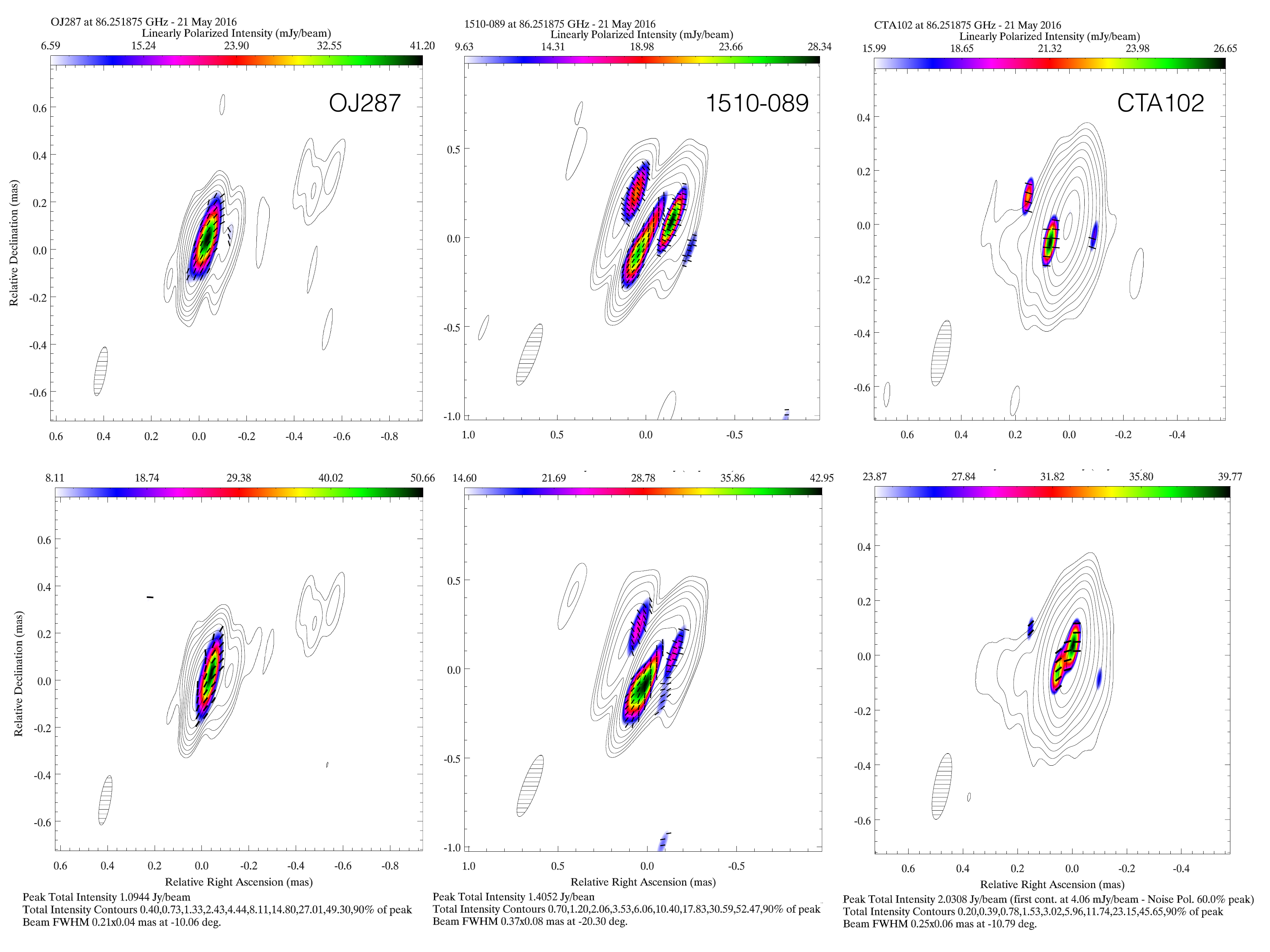}
\caption{This figure shows 3 mm GMVA images in total (contours) and linearly polarized (colours) intensity  of OJ~287, 1510-089, and CTA~102 taken on 21 May 2016. Black sticks represent the electric vector position angles. The upper panels are the resulting images after applying to the sources their own {\textit D-terms}, while in the lower panel we applied the average {\textit D-terms} to all the sources.}
\label{pol_test}
\end{figure}

From the comparison of the linearly polarized images obtained applying the two methods for the calibration of the instrumental polarization, we
observe that, in general, the images obtained present some differences. In
particular, we found that only in the case of OJ~287 (best case) is the
morphology of the polarized flux similar in both images, as can be seen in
Figure \ref{pol_test}. On the other hand, for the other two sources (1510-089 and CTA~102),
the morphology of the polarized flux changes substantially between the two images, and the peak of the linearly polarized flux shifted to a different position when we applied their own {\textit D-terms} (Figure \ref{pol_test}, upper panels).
Moreover, we notice that for all three sources, the polarized flux decreased when we applied their own {\textit D-terms}.

Since many sources from the sample have a parallactic angle coverage comparable to those of 1510-089 and CTA~102 (less than 100$^{\circ}$ for most of the antennas), and certainly worse than that of OJ~287, plus some sources also have more complex polarized substructures, we decided to~calibrate the instrumental polarization using the average {\textit D-terms}.
This approach has the significant advantage of obtaining the final {\textit D-term} values to apply by taking into account a large number of source measurements and not only one. This is of particular importance for a mm-VLBI dataset, where~the choice of a calibrator for the polarization is not trivial and also depends on the quality of~data.

\begin{figure}[H]
\centering
\includegraphics[width=13 cm]{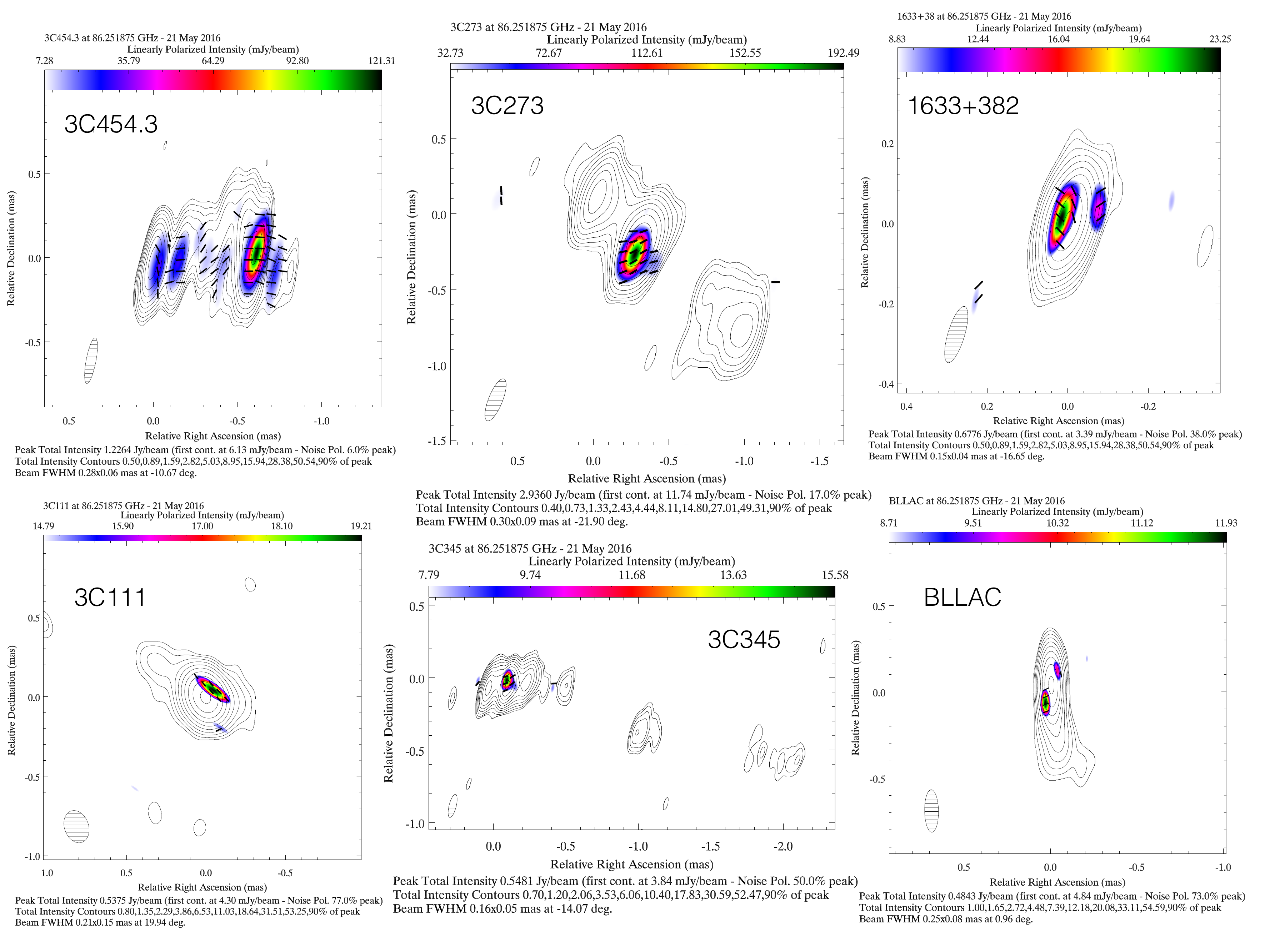}
\caption{This figure shows 3 mm GMVA images of the quasars 3C~454.3, 3C~273, 1633+382, 3C~345, the radio galaxy 3C~111, and BL Lac. Colours, contours, and black sticks represent the same as in Figure \ref{pol_test}.}
\label{pol_images}
\end{figure}

Some of the final linearly polarized (in colors) and
total intensity (in contours) GMVA images are shown in Figure \ref{pol_images}, where the absolute calibration of EVPAs has also been applied.~The rms reached in all the total intensity images was $\sim$1--3 mJy/beam and in polarized images was $\sim$4--7 mJy/beam. The~median value of the beam considering all images was $\sim$0.2 $\times$ 0.06 mas.

\vspace{-6pt}

The high resolution reached with these mm-VLBI observations allows us to
distinguish features not visible in 43 GHz VLBA observations (e.g., the polarized
features at $\sim$0.1 mas from the core in 3C~454.3 and 3C~345). These features become apparent when we
compare the GMVA data presented here with data from the VLBA-BU-BLAZAR program of a nearby epoch (10 June 2016), available at the program's webpage. A possible complex structure of the 3C 454.3 core was discussed in \cite{Jorstad:2010uq} using super-resolved images at 7 mm.
Another quite noticeable result from the GMVA observations is seen in the image of the quasar
1510-089, where at 86 GHz we see an unusual feature on the
eastern side of the source, here distinguishable in both total and linearly polarized
intensity maps but visible only in the polarized flux in the 43 GHz VLBA data.

Moreover, a comparison between the fractional polarization in GMVA and 43 GHz VLBA data reveals slightly higher values in GMVA data (between 2$\%$ and 9$\%$) than in 43 GHz VLBA data  (between 1$\%$ and 6$\%$), as expected because of lower beam depolarization at shorter wavelength.

%%%%%%%%%%%%%%%%%%%%%%%%%%%%%%%%%%%%%%%%%%
\section{Conclusions}

We have presented the most complete sample so far of 3 mm GMVA polarized images of AGN jets.
We showed that 3 mm GMVA observations are a powerful tool to investigate the
central regions of distant blazars and radio galaxies, thanks to the reduced
opacity at 3 mm and the improved angular resolution that allow us to distinguish
features not visible in 43 GHz VLBA observations (e.g., 3C~454.3, 3C345, and
1510-089).

From the comparison of two different methods for the calibration of the instrumental
polarization, we tested that the {\textit D-terms} of a source are well-defined
only in the case of good coverage of the parallactic angle, as expected. Moreover, we
found that a poor parallactic angle coverage can lead to changes in the source polarized structure and intensity.

The use of average {\textit D-terms} instead of the {\textit D-terms} proper of
the source itself for the calibration of instrumental polarization turns out to
be a more stable method, also producing more reliable polarized intensity images.

%%%%%%%%%%%%%%%%%%%%%%%%%%%%%%%%%%%%%%%%%%
\vspace{6pt}

%%%%%%%%%%%%%%%%%%%%%%%%%%%%%%%%%%%%%%%%%%
\acknowledgments{This research has made use of data obtained with the Global Millimeter VLBI Array (GMVA),
which consists of telescopes operated by the MPIfR, IRAM, Onsala, Metsahovi, Yebes and the
VLBA. The data were correlated at the correlator of the MPIfR in Bonn, Germany.~The VLBA
is an instrument of the Long Baseline Observatory. The Long Baseline Observatory is a
facility of the National Science Foundation operated by Associated Universities, Inc.( Washington, DC, USA)
The research at Boston University was supported by NASA through a number of Fermi Guest Investigator program grants, most recently NNX14AQ58G.
IRAM 30~m Telescope is supported by INSU/CNRS (France), MPG (Germany) and IGN (Spain). IA acknowledges support by a Ram\'on y Cajal grant of the Ministerio de Econom\'ia, Industria y Competitividad (MINECO) of Spain. The research at the IAA-CSIC was supported in part by the MINECO through grants AYA2016--80889--P, AYA2013-40825-P, and AYA2010-14844, and by the regional government of Andaluc\'{i}a through grant P09-FQM-4784. J.-Y. Kim is supported for this research by the International Max-Planck Research School (IMPRS) for Astronomy and Astrophysics at the University of Bonn and Cologne.
We would also like to thank the MPIfR internal referee B. Boccardi.}

\reftitle{References}

\end{document}